\documentclass[pra,final,twocolumn,showpacs,showkeys]{revtex4}
\usepackage{amsmath}
\usepackage{graphicx}
\usepackage{amsfonts}
\usepackage{amssymb}
\usepackage{subfigure}

\graphicspath{ {./imgs/} {./asymptote/} }

\DeclareMathOperator{\ham}{\mathcal{H}}
\DeclareMathOperator{\hcol}{\mathcal{H}_{col}}

\newcommand{\eps}{\epsilon}

\newcommand{\bra}[1]{\langle\, #1 \,|}
\newcommand{\ket}[1]{|\, #1 \,\rangle}
\newcommand{\braket}[2]{\langle\,#1\,|\,#2\,\rangle}
\newcommand{\bc}[1]{\bra{\mbox{col} \; #1}}
\newcommand{\kc}[1]{\ket{\mbox{col} \; #1}}

\newcommand{\subs}[1]
{ 
	\mbox{\scriptsize{#1}}
}

\begin{document}

\title{Quantum Walks on Trees with Disorder: Decay, Diffusion, and Localization}

\author{Steven R. Jackson}
\author{Teng Jian Khoo}
\author{Frederick W. Strauch}
\email[Electronic address: ]{Frederick.W.Strauch@williams.edu}
\affiliation{Williams College, Williamstown, Massachusetts 01267}

\date{\today}

\begin{abstract}
Quantum walks have been shown to have impressive transport properties compared to classical random walks.  However, imperfections in the quantum walk algorithm can destroy any quantum mechanical speed-up due to Anderson localization.  We numerically study the effect of static disorder on a quantum walk on the glued trees graph.  For small disorder, we find that the dominant effect is a type of quantum decay, and not quantum localization.  For intermediate disorder, there is a crossover to diffusive transport, while a localization transition is observed at large disorder, in agreement with Anderson localization on the Cayley tree.  

\end{abstract} 
\maketitle

\section{Introduction}

Quantum walks constitute a promising route to the development of algorithms for quantum information processing.  
Successful applications of quantum walks include a variety of search algorithms \cite{Shenvi2003,Childs2004a,Ambainis2004a,Childs2004b}, graph hitting problems \cite{Childs2002,Childs2003,Kempe03a}, Boolean function evaluation \cite{Farhi2008, Childs2009a}, among others. Quantum walks are in fact a universal paradigm for quantum computation \cite{Childs2009, Lovett2010}.  

\begin{figure}[t]
\begin{center}
\includegraphics[width=3.5in]{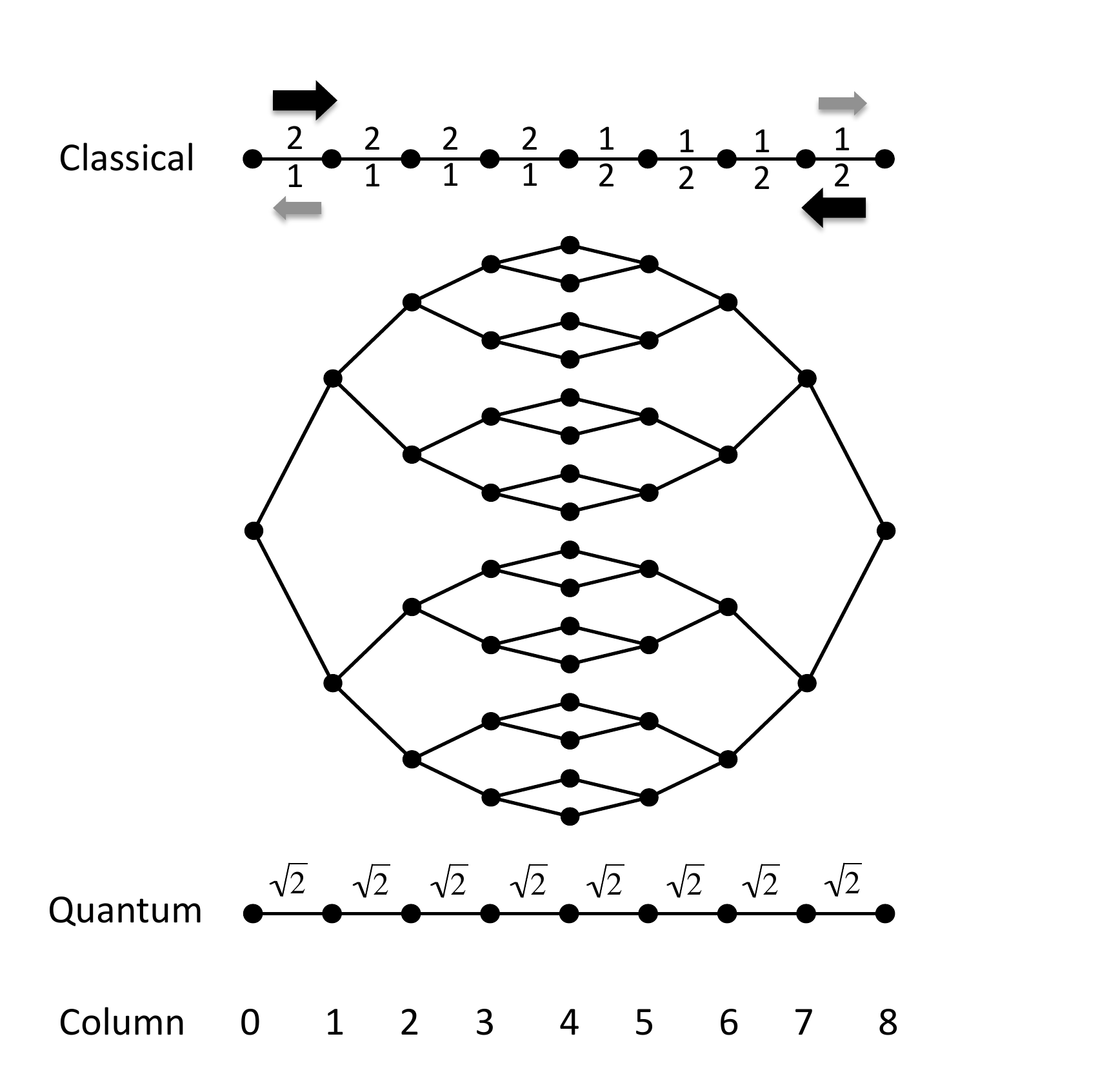}
\caption{Glued binary trees graph $G_4$, showing the reduction to the column space for both the classical (top) and quantum (bottom) random walks. }
\label{fig1}
\end{center}
\end{figure}

The simplest and most dramatic improvement demonstrated by quantum walks, when compared with the corresponding classical random walks, is the hitting probability of a walk on certain graphs.   Two graphs in particular have demonstrated an exponential separation between the quantum and classical walks: the glued binary trees \cite{Childs2002} and the hypercube \cite{ Kempe03a}.  An example of the former is shown in Fig. \ref{fig1}; this graph is formed by connecting the leaves of two binary trees of depth $d$.  Due to the graph's symmetry, the quantum walk can be restricted to a subspace of the total (exponentially large) Hilbert space, known as the ``column space'' (also shown in Fig. 1).  It has been argued that this symmetry is the heart of the quantum speed-up \cite{Krovi2007}.  A quantum walk algorithm on a slight modification of this graph, to be described below, was proven to have an exponential speedup over any classical algorithm \cite{Childs2003}.

Keating {{\it et al.}} have argued that physical implementations of these walks will not be able to maintain this quantum speed-up for large graphs \cite{Keating07}.   Indeed, one expects that physical systems will be subject to decoherence and disorder.  While one can always argue that future quantum computers will be protected from these effects by error correction, it is more likely that near term demonstrations of quantum walks will use a physical network.  In fact, many of the recent experiments on quantum walks involve an encoding of the degrees of freedom that is not, strictly speaking, computationally useful \cite{Kendon11}.   Nevertheless, these experiments demonstrate the dynamical speed-up characteristic of  quantum walks.  Determining when disorder or decoherence will limit the observability of this speed-up is an important theoretical problem \cite{Kendon07}.

This problem may already have been encountered by nature.  Recent evidence has shown that photosynthetic complexes operate using a type of quantum walk \cite{AGuzik08}, where the interplay of decoherence with disorder plays a crucial role in their energy harvesting efficiency.   This phenomenon, known as dephasing or environmentally assisted transport, has relations to earlier studies of phonon effects on electron transport in disordered solids \cite{Jonson79}.  Understanding and potentially reverse-engineering this efficiency is a topic of great scientific and practical importance.  As the complexes and potential technologies under study can be far from the thermodynamic limit, examining quantum transport on graphs of modest size may reveal new surprises.

In this paper, we examine the continuous-time quantum walk on the glued trees graph with static disorder, corresponding to the usual Anderson model with disordered on-site energies \cite{Anderson58}.  Previous authors \cite{Keating07} introduced disorder through an effective one-dimensional representation (the column space).  Using the fact that all eigenstates are localized in one-dimension with arbitrary disorder, they concluded that Anderson localization would cause an exponential suppression of the hitting probability.  From a physical perspective, however, this model has some flaws.  First, their model introduces a highly correlated form of disorder, in that all of the on-site energies on a given column are the same  Second, the glued trees graph is quite similar to a Cayley tree, whose dimensionality is formally infinite, as far as Anderson localization is concerned \cite{Evers2008}.  The Cayley tree exhibits a localization transition at large disorder, and thus significant speed-up may still be possible for weak disorder.  

We have performed a numerical study of this problem for the full Hilbert space of the glued trees graph and modifications thereof.  Analysis of the eigenvalues and eigenvectors agree with previous studies of localization the Cayley tree.  We pay special attention to the less well studied regime of weak disorder, using dynamical simulations to find a type of quantum decay of the quantum walk from the column space.  The resulting hitting probability is well simulated by a model with local decay from each column state.   We further consider the crossover from wavelike to diffusive transport for intermediate levels of disorder.  A scattering theory analysis of this regime indicates a transition from the quantum walk to a classical random walk. 

Our results augment the many results for one-dimensional quantum walks with disorder.  These include theoretical studies for the continuous-time quantum walk \cite{Yin2008} and the discrete-time quantum walk \cite{Ahlbrecht2011}, and the many recent experiments on atoms \cite{Karski2009}, ions \cite{Schmitz2009,Zahringer2010}, and photons \cite{Broome2010, Schreiber2011}.   Higher-dimensional quantum walks could be realized in these or other systems, such as networks of superconducting circuits  \cite{Strauch2008, Chudzicki10}.   In particular, our simulations show that the effects of disorder on quantum transport can be identified in systems of 10-100 lattice sites.   

This paper is organized as follows.  In Section II we review the known results for Anderson localization on the Cayley tree, and in Section III numerically study the phase diagram for the two types of glued trees, and find evidence of the localization transition.  In Section IV we study the time-dependence of the quantum walk with disorder, and introduce the local decay model for weak disorder.  In Section V we consider a scattering theory model for transport through the glued trees graph, and find evidence for the transition to classical random walk.  We conclude in Section VI, comparing our results with the hypercube and highlighting the major open questions.  Results for the quantum scattering and (classical) diffusive transport through the glued trees are found in the Appendices.

\section{Localization and Diffusion on the Binary Tree}

Anderson localization is the phenomenon that, for a sufficiently large amount of disorder, the eigenstates of a quantum system become exponentially localized about the nodes in a graph \cite{Anderson58}.  Localization transitions are a rich field of study \cite{Evers2008}, for which both symmetry and dimensionality play key roles.   The quantum walk we consider corresponds to the simplest such model, a tight-binding Hamiltonian with random on-site energies
\begin{equation}
\mathcal{H} = - \gamma \sum_{\langle j,k \rangle} \left(c_j^{\dagger} c_k + c_k^{\dagger} c_j \right) + \sum_{j} \eps_j c_j^{\dagger} c_j,
\label{andersonmodel}
\end{equation}
where $\gamma$ is a hopping rate, $\eps_j$ is a set of random variables, uniformly distributed in the range $[-W/2,W/2]$, and $c_j^{\dagger}$ is the creation operator for an excitation at site $j$, and the sum is over neighboring sites.  This model, originally inspired by electron transport in solids, can be used for many physical systems, such as quantum state transfer of a single excitation on a qubit network \cite{Bose2003,Bose2008,Strauch2008} or exciton transfer in a photosynthetic complex \cite{AGuzik08}.  

Much is known about the Anderson model given by Eq. \ref{andersonmodel} \cite{Evers2008}.  For example, for a one-dimensional infinite lattice, localization occurs for arbitrarily small amounts of disorder \cite{Borland63}.  It was argued that this property was generic to quantum walks \cite{Keating07}.  However, it is known that for systems with dimension greater than two, localization requires a sufficiently large amount of disorder, i.e. there is a localization transition as the disorder is increased \cite{Abrahams79}.   At first glance, the glued trees graph of Fig. 1 might appear to be a subset of a two-dimensional system, and thus one might expect localization for even small amounts of disorder.

In fact, the infinite binary tree, or more generally the infinite Cayley tree (also known as the Bethe lattice \cite{Ostilli12}) has been used as a model for an infinite-dimensional system.  This has been studied extensively, and exhibits a localization transition found numerically \cite {Jonson79,Miller94,Monthus2009,Biroli2010} and by an analytical mean field calculation \cite{AbC73,AbC74}.  For this system, there is a mobility edge in the energy spectrum, such that eigenstates inside the mobility edge are extended, while those outside are localized.   For sufficiently large values of disorder, there is a localization transition beyond which all eigenstates are localized.  For a binary tree (or Cayley tree with $K=1$), this transition occurs for $W_c \approx 17$ \cite{AbC74,Jonson79,Miller94,Monthus2009,Biroli2010}.  

The description of localization given above, in which the eigenstates of the system exhibit a transition from extended to localized states, can be called eigenstate localization.  There are two other ways to identify localization that we will encounter in this paper.  The first is dynamical localization, in which the spreading of a wavepacket shows a saturation as a function of time.  The second is spectral localization, in which the eigenvalues of the system move from an absolutely continuous spectrum (corresponding to extended states) to a pure point spectrum (corresponding to localized states).  These other indicators of localization, which have also been studied extensively, we now briefly summarize.  

The spectral properties of the system were studied numerically \cite{Sade2003,Biroli2010}, and found to exhibit a transition in agreement with the self-consistent approaches described above.  These numerical studies are sensitive to the handling of the boundary of the tree \cite{Aizenman2006}, a fact to which we will return in Section III.  While there are some subtleties regarding the the phase diagram at weak disorder \cite{Aizenman2011a,Aizenman2011b}, the existence of a localization transition for large disorder is well established.  For small disorder, the existence of extended eigenstates has been proven \cite{Klein98}.  It has also been proven that, for small disorder,  states that are initially localized spread ballistically \cite{Klein95}.  

The ballistic spreading does not preclude classical behavior, however.  The random walk on the Bethe lattice has been studied, and can be mapped onto an asymmetric random walk on a one-dimensional half-line \cite{Hughes82}, as indicated in Fig. \ref{fig1} ({\it e.g.} the left half).  A classical walker is twice as likely to move right as left, and this asymmetry leads to the peculiar fact that the classical walk also spreads ballistically.  Exact results and limits have been established for this and other properties of the classical walk \cite{Cassi89, Monthus96}.  Note that the asymmetry is linked to the exponential growth of sites away from the origin, such that one often calls the Bethe lattice a graph of infinite dimensionality.  

\section{Quantum Walk Eigenstates}

The continuous-time quantum walk \cite{Farhi98,Mulken2011} is precisely the quantum dynamics of a single particle moving on a graph.  This is given by the Schr{\"o}dinger equation
\begin{equation}
i \frac{d \psi_j}{d t} = - \gamma \sum_{k} A_{jk} \psi_k,
\end{equation}
where $A_{jk}$ is the adjacency matrix for the graph and $\gamma$ is hopping rate.  One could also use the Laplacian matrix, or introduce potentials to implement search algorithms \cite{Childs2004a,Childs2004b}, but here we consider the addition of static disorder, such that  
\begin{equation}
i \frac{d \psi_j}{d t} = - \gamma \sum_{k} A_{jk} \psi_k + \eps_j \psi_j,
\end{equation}
where the on-site energies $\eps_j$ are i.i.d. random variables uniformly distributed in the range $[-W/2,W/2]$.  

In this section we consider the nature of the eigenstates of the Hamiltonian of the quantum walk with disorder, namely $\mathcal{H} |\Psi\rangle = E |\Psi\rangle$, with $\mathcal{H} = \mathcal{H}_0 + \mathcal{H}'$, with the unperturbed Hamiltonian given by
\begin{equation}
\mathcal{H}_0 = -\gamma \sum_{j,k=1}^{N_d} A_{jk} |j\rangle \langle k|,
\end{equation}
and diagonal static disorder
\begin{equation}
\mathcal{H}' = \sum_{j=1}^{N_d} \eps_j |j\rangle \langle j|,
\label{hprime}
\end{equation} 
where $N_d = 3 \times 2^d -2$ is the number of vertices for the glued trees graph $G_d$.  We begin by analyzing $\mathcal{H}_0$, and follow by studying the eigenvalues and eigenstates of $\mathcal{H}$.

\subsection{Eigenstates without disorder}

As described above, there is a great deal of symmetry in the glued trees graph $G_d$, and a great deal of structure in the eigenstates and eigenvalues of the system.  We begin by providing a notation for the graph.  We consider a labeling along the ``columns'' and ``rows'' of the graph, of the form $(j,n)$, where $j = 0 \to 2d$ indicates the distance from the left root along the graph, and $n = 0 \to N_{j,d}-1$ is the location within a given column. Here $N_{j,d}$ is the number of sites in a given column $j$,  given by $N_{j,d} = 2^j$ for $j \le d$, and $N_{j,d} = 2^{2d-j}$ for $j > d$.  The coordinates $(j,n)$ can be combined into a single coordinate $v$ by the following rule
\begin{equation}
v = \left\{ \begin{array}{ll} 
2^j + n & \mbox{for} \ 0 \le j \le d, \\
3 \times 2^d - 2^{2d + 1 - j} + n & \mbox{for} \ d < j \le 2 d.
\end{array} \right.
\end{equation}
Note that $v$ ranges from $1 \to 3 \times 2^d -2$.  

The adjacency matrix elements $A_{v,w}$ are equal to one if vertices $v$ and $w$ are connected, and zero otherwise.  This can be given in terms of the coordinates $(j,n)$ by observing that $(j,n)$ is connected to
\begin{equation}
(j-1, \lfloor n/2 \rfloor), (j+1,2n), (j+1,2n+1) \ \mbox{for} \ j \le d,
\end{equation}
and
\begin{equation}
(j+1, \lfloor n/2 \rfloor), (j-1,2n), (j-1,2n+1) \ \mbox{for} \ j > d.
\end{equation}
What is most important is that a given vertex is symmetrically coupled to those sites one step further from the left root (or closer to the right root).  This allows us to express the eigenstates of the system in terms of ``column-states'' that are equal superpositions of states on a given column $j$.  This column-space reduction is well-known for its utility in the analysis of the quantum walk \cite{Childs2002,Childs2003, Krovi2007}.

Letting $|j,n\rangle$ denote the Hilbert-space vector associated with vertex $(j,n)$, we define the column-space vector $|\mbox{col} \ j \rangle$ by
\begin{equation}
|\mbox{col} \ j \rangle = \frac{1}{\sqrt{N_{j,d}}} \sum_{n=0}^{N_{j,d}-1} |j,n\rangle.
\end{equation}
By the symmetry noted above, the Hamiltonian $\mathcal{H}_0 = -\gamma A$ acts on the column-space states as
\begin{equation}
\mathcal{H}_0 |\mbox{col} \ j \rangle = -\sqrt{2} \gamma | \mbox{col} \ j-1 \rangle - \sqrt{2} \gamma | \mbox{col} \ j+1\rangle,
\end{equation}
where one factor of $\sqrt{2}$ is due to the number of connections to a neighboring column, and the other due to the normalization of the column states.  Hence, we can reduce the dynamics to a quantum walk on a finite line, whose eigenstates are equally well-known
\begin{equation}
	\ket{\Psi_{k,d}} = \frac{1}{\sqrt{d+1}}\sum_{j=0}^{2d} \,\sin\!\left(\frac{k\,(j+1)\,\pi}{2(d+1)}\right)\ket{ \mbox{col} \ j},
	\label{ColSpEs}
\end{equation}
with energies
\begin{equation}
	E_{k,d} = - 2\sqrt{2}\,\gamma \,\cos\!\left(\frac{k\,\pi}{2(d+1)}\right), \label{ColspEn}
\end{equation}
and $k = 1 \to 2d+1$.  We have annotated the states by the depth of the graph, as there are in fact many more eigenstates for $G_d$, whose enumeration we now consider.  

\begin{figure}[t]
\begin{center}
\includegraphics[width=3in]{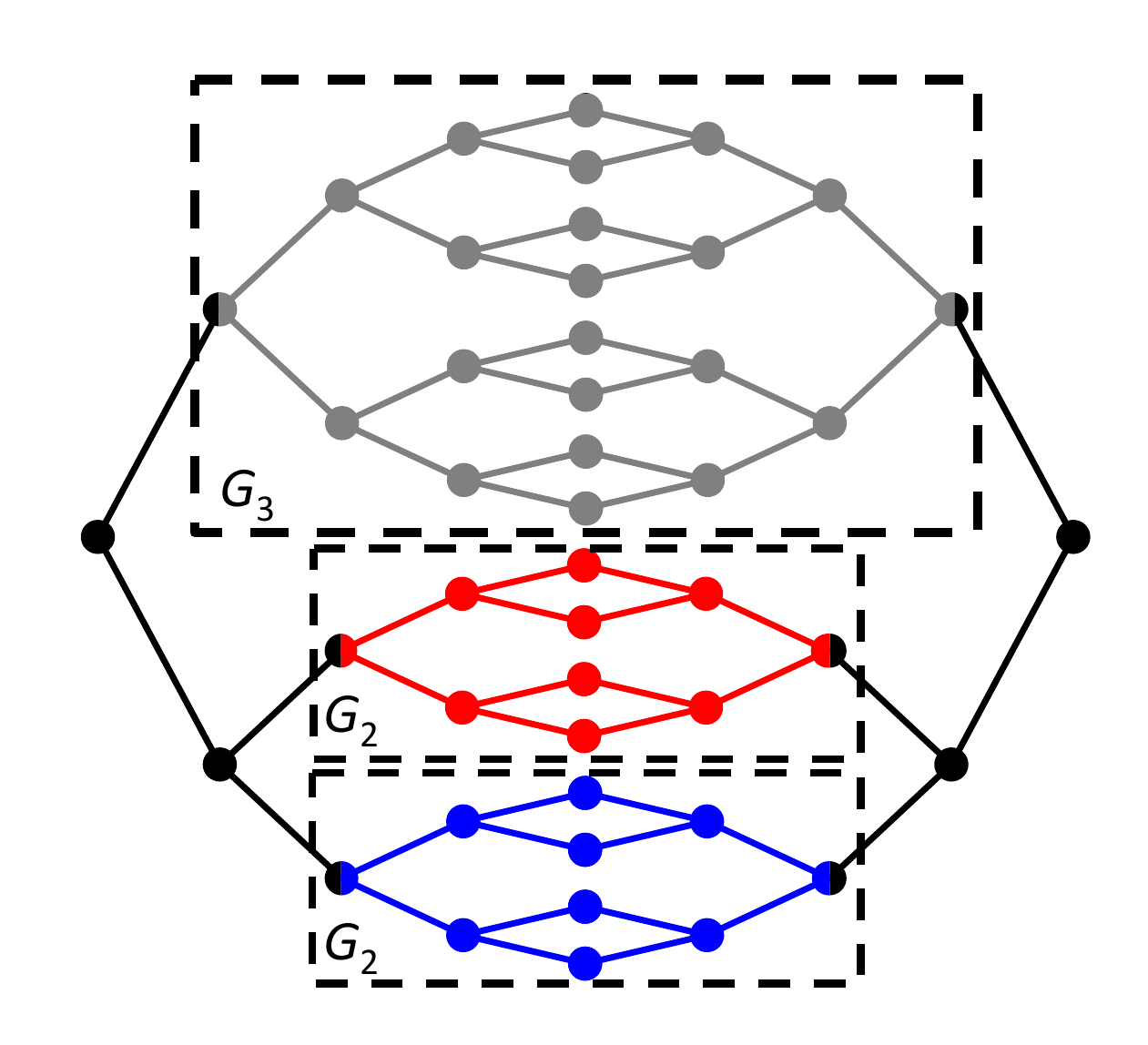}
\caption{(Color online) Glued binary trees graph $G_4$, with several highlighted subgraphs $G_3$ (top, in gray) and two copies of $G_2$ (middle, in red and bottom, in blue). }
\label{subgraphs}
\end{center}
\end{figure}

The glued trees graph is self-similar, in that $G_d$ contains $2^\nu$ subgraphs, 
each equivalent to $G_{d-\nu}$.  These subgraphs can be grouped into $2^{\nu-1}$ pairs, each pair formed by removing the roots of a larger graph equivalent to $G_{d-\nu+1}$.  That is, we can repeatedly split the glued trees graph by removing the left and right roots, such that $G_d$ contains 2 copies of $G_{d-1}$, 4 copies of $G_{d-2}$, and so forth, formed by removing the roots until we are left with $2^d$ copies of $G_0$ (the isolated vertices at the center of the graph).  Some representative subgraphs of $G_4$ are shown in Fig. \ref{subgraphs}.  On their own, each subgraph would have eigenstates of the form of Eq. (\ref{ColSpEs}).  To find how these subgraphs contribute to the spectrum of $G_d$, we observe that an equal but opposite-signed superposition over two paired subgraph eigenstates is an eigenstate of $G_d$.  This occurs because the components with opposite phase, when acted upon by $\mathcal{H}_0$, will interfere destructively on the two nodes to which they are connected on the larger graph.  By this pairing, we can thus construct the complete set of eigenstates for $G_d$.   

To see this more clearly, we define a set of ``sub-column'' states whose elements combine the paired subgraphs described above.  These are given by
\begin{equation}
|\mbox{scol} \ j; \alpha, \nu \rangle =  \sum_{n= 2 \alpha N_{d-\nu}}^{ (2 \alpha +1) N_{d-\nu} - 1 } \frac{ |j+\nu,n\rangle - |j+\nu, n + N_{j,d-\nu} \rangle}{\sqrt{2 N_{j,d-\nu}}},
\end{equation}
where $j = 0 \to 2(d-\nu)$ indicates the column in $G_{d-\nu}$, $\alpha = 0 \to 2^{\nu-1}-1$ labels the paired subgraphs, and $\nu = 1 \to d$ indicates the depth of the subgraphs' left root.  Using these states, the remaining eigenstates of $G_d$ can be obtained by using the eigenstates $|\Psi_{k,d-\nu}\rangle$ from Eq. (\ref{ColSpEs}) with $|\mbox{col} \ j \rangle$ replaced by $|\mbox{scol} \ j; \alpha, \nu\rangle$ and eigenvalues $E_{k,\,d-\nu}$ from Eq. (\ref{ColspEn}), where $k = 1 \to 2(d-\nu)+1$.  Defining
\begin{equation}
\sigma_d = \{ E_{k,d}, k = 1 \to 2 d \},
\end{equation}
the total spectrum can then be written as
\begin{equation}
\sigma = \sigma_d + \sum_{\nu=1}^{d} 2^{\nu-1} \sigma_{d-\nu}.
\end{equation}
This spectrum exhibits a very large degeneracy, especially for $E=0$, which has a multiplicity of $2^d$ (one from each copy of $\sigma_{d-\nu}$, and half from the $\sigma_0$).  This large degeneracy (also observed in \cite{Aizenman2006}) occurs for trees that are both glued and unglued and can make numerical analysis of the disordered system problematic, as will be described below.  

\subsection{Eigenstates with disorder}

The introduction of static disorder changes both the eigenvalues and eigenvectors of the system.  For the Cayley tree, early studies used a self-consistent approach \cite{AbC74,Miller94} to find the localization transition and a phase diagram between extended and localized eigenstates.  Recent numerical studies \cite{Monthus2009,Biroli2010} have confirmed these earlier results, which we now summarize.

The eigenvalue spectrum has been studied for a single Cayley tree numerically by \cite{Sade2003} through the use of spectral statistics.  A transition of the distribution of the energy level spacings from Wigner to Poisson (indicative of a localization transition) was observed when the tree was modified so that the leaves are randomly connected to each other.  This random connection presumably reduces the prevalence of the zero eigenvalue described above, which would otherwise lead to a Poisson distribution for the spectral statistics \cite{Aizenman2006}.  We performed a similar analysis for the glued trees graph, with the leaves connected as in Fig. \ref{fig1}, which we call a simple glued trees (SGT) graph, or interrupted by a random cycle (as in \cite{Childs2003}), which we call a modified glued trees (MGT) graph.  An example of the MGT (with a regular connection \cite{Douglas09}) is shown in Fig. 9.  A Wigner to Poisson transition was observed for the MGT, while the SGT exhibited a spectrum that was always far from Wigner.  For the remainder of this section, our results were obtained using the MGT.  
 
The localization of the eigenstates and the localization transition can be found by studying a simple property of the eigenstates, namely the inverse participation ratio (IPR) $I_2$:
\begin{equation}
I_2(\psi) = \sum_{j} |\psi_j|^4,
\end{equation}
where we assume that the states are normalized with $I_1(\psi) = \sum_{j} |\psi_j|^2 = 1$.  This quantity has the property that an eigenstate localized to a single site has $I_2(\psi) = 1$, while an eigenstate extended over $N$ sites has $I_2(\psi) = 1/N$.  We further define an  averaged value of this quantity
\begin{equation}
I_2(E) = \frac{1}{N(E;\Delta E)} \sum_{  |\langle \psi | \ham | \psi \rangle - E| < \Delta E} I_2(\psi),
\end{equation}
where $N(E;\Delta E)$ is the number of eigenstates found to have eigenvalue $E_j$ with  in the range $E-\Delta E < E_j < E+\Delta E$, leaving the dependence on $\Delta E$ implicit..  By calculating $I_2(E)$ the eigenstates of the system as a function of energy and disorder, the phase diagram and localization transition can be visualized.

\begin{figure}
\begin{center}
 \includegraphics[width = 3.5in]{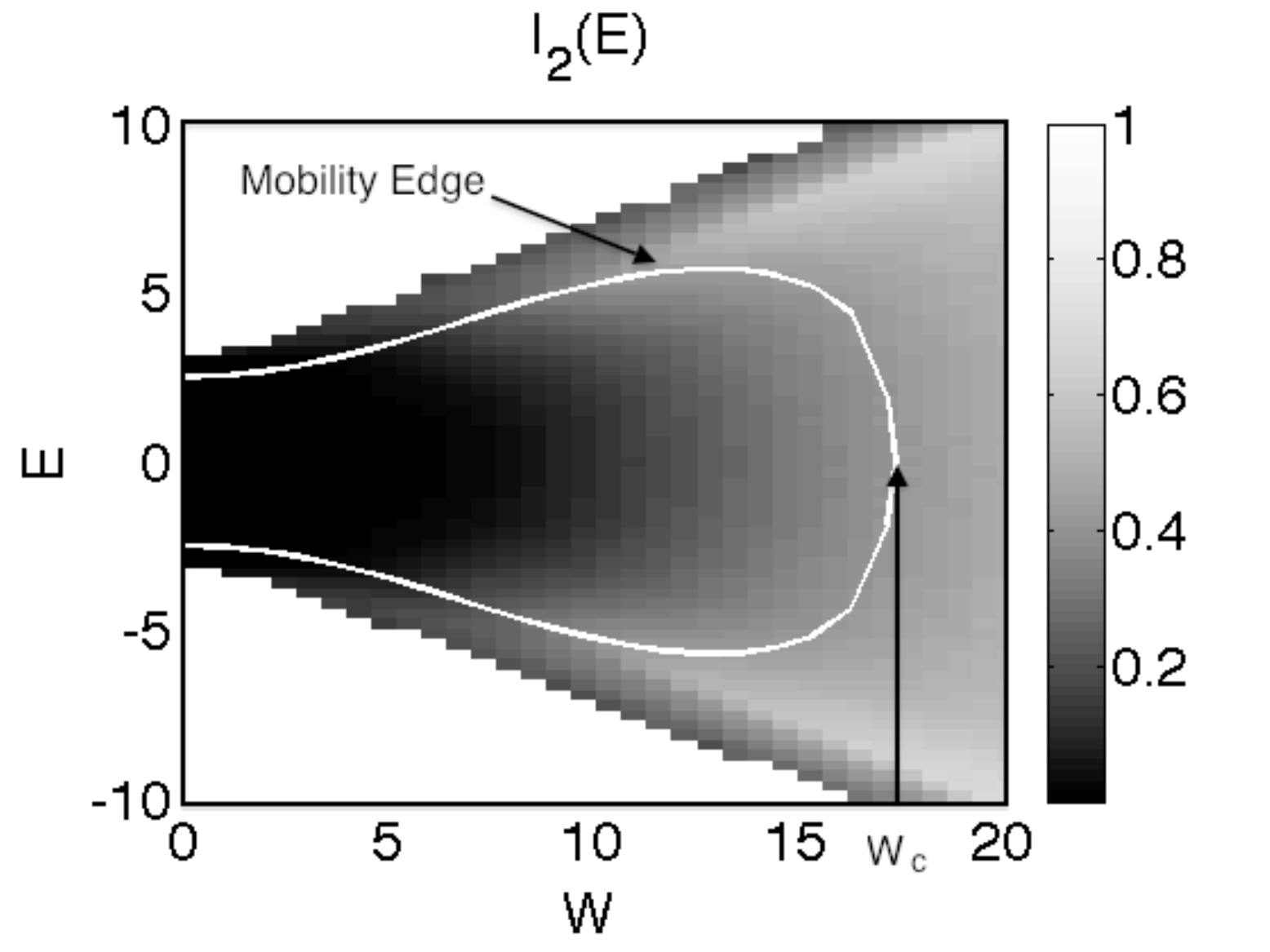}
\caption{Inverse participation ratio $I_2(E)$ as a function of energy and disorder, for $d=8$.  For each value of disorder the IPR was averaged over 500 realizations of $\mathcal{H}$ with $\Delta E = 0.15$ (with $\gamma = 1$).  Also shown is the mobility edge from the self-consistent theory, predicting a localization transition with $W_c \approx 17$.}
\label{IPRfig1}
\end{center}
\end{figure}

For the SGT, we again encounter a difficulty in that there are a large number of states with an $I_2(\psi) = 1/2$, associated with the $2^{d-1}$ states in $\sigma_0$ (with $E=0$).  However, by using the MGT, we can calculate meaningful results for $I_2(E)$ for various disorder strengths $W$ to obtain the diagram in \ref{IPRfig1}.  Here we have let $d = 8$ and set $\Delta E = 0.15 \gamma$ and averaged over 500 realizations of $\mathcal{H}$.   We observe a gradual movement of extended states (with small $I_2$) to localized states (with $I_2 > 1/2$) as disorder is increased.  Also shown are the expected results for an infinite Bethe lattice, obtained using the self-consistent method of Miller and Derrida \cite{Miller94}  

To obtain an estimate of the localization transition, we fix our attention to states near $E=0$, and repeat the calculation of the averaged IPR for graphs of various sizes.  As expected, $I_2(0)$ exhibits a small size-dependent value for $W=0$, which increases to approximately 0.5 for large $W$.  At a certain value, the averaged IPRs for all of the graphs coalesce, from which we estimate $W_c \approx 17$, in agreement with recent results for the Cayley tree \cite{Monthus2009,Biroli2010}.

\begin{figure}
\begin{center}
 \includegraphics[width = 3.5in]{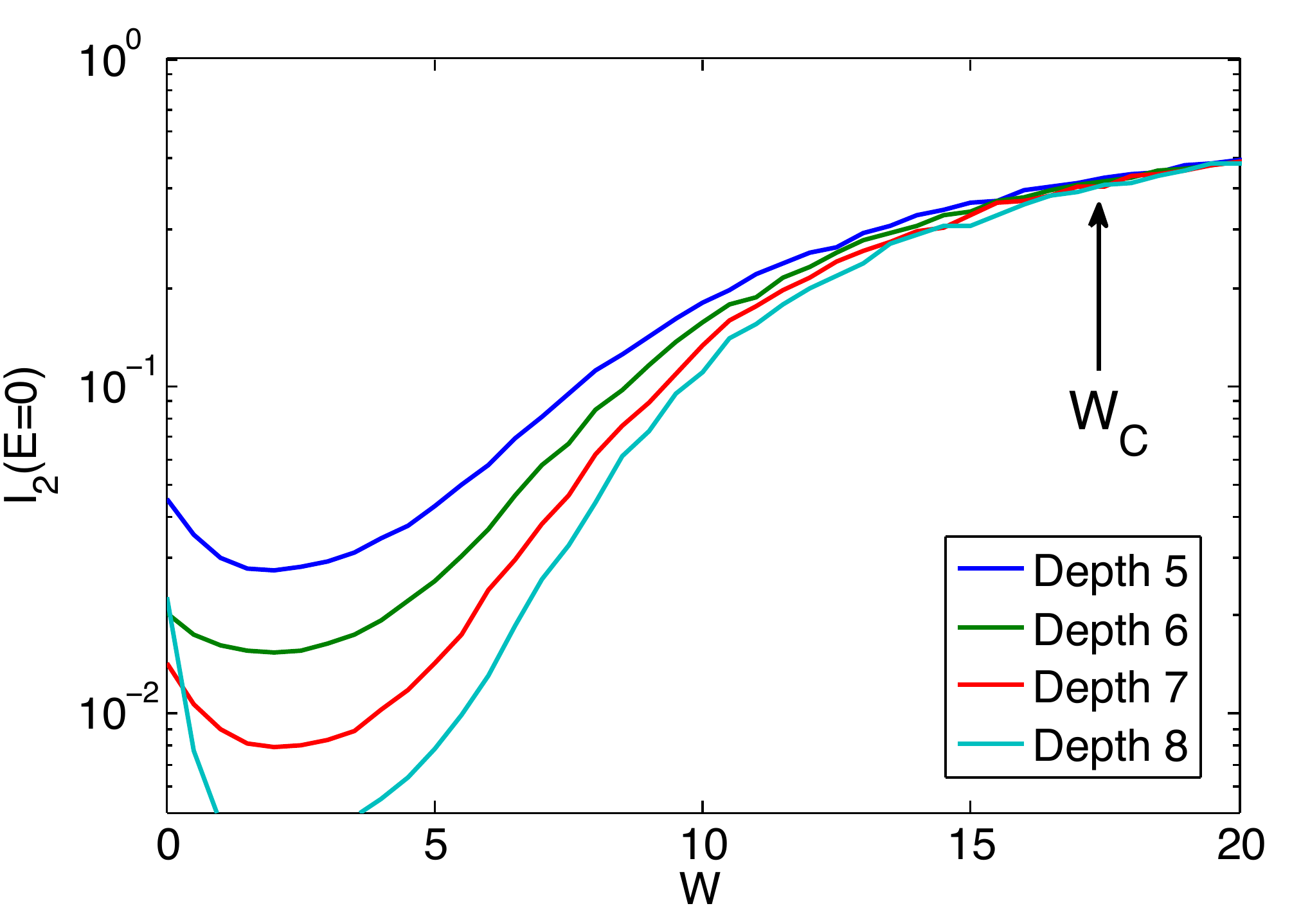}
\caption{(Color online) Inverse participation ratio $I_2(E)$ at the band center ($E=0$) as a function of disorder for various depths.  For each value of disorder the IPR was averaged over the middle 100 eigenvalues and a number of realizations given by $500, 250, 125, \mbox{and} \ 50$ for $d = 5, 6, 7, \mbox{and} \ 8$, respectively.  The localization transition occurs when the curves collapse near $W \approx 17$.}
\label{IPRfig2}
\end{center}
\end{figure}

\section{Quantum Walk Dynamics}

The dynamics of a quantum walk on the glued trees graph with disorder has been studied using the one-dimensional column-space model by Keating {{\it et al.}} \cite{Keating07}.   In the previous section, however, we have seen that the eigenstates of the MGT undergo a localization transition at large disorder.  This leaves open the possibility that a dynamical speed-up can be observed for small disorder.   To explore this possibility, we again turn to numerical studies.  There are three issues to be studied: first, what is the probability to hit the right root should the quantum walk begin at the left root?  Second, how does this probability decay as a function of disorder and size, and by what mechanism?  Finally, if the disorder quantum walk does not hit the right root, how far does it get?

\begin{figure}[t]
\begin{center}
\includegraphics[width=3.5in]{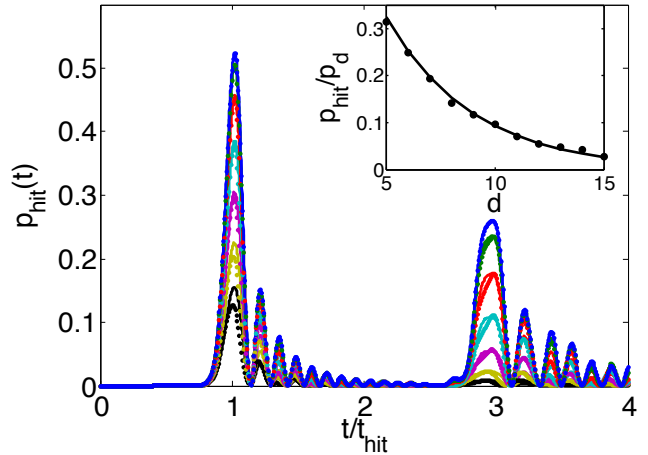}
\caption{(Color online) Hitting probability $p_{\subs{hit}}(t)$ as a function of time for the quantum walk on the glued trees graph with $d=15$ and various values of disorder.  The symbols were calculated using exact numerical simulations with (from top to bottom) $W = 0, 0.2, 0.4, 0.6, 0.8, 1.0, \mbox{and} \ 1.2$, with 10 realizations of $\mathcal{H}$ for each value of $W$.  The solid curves were calculated using the locay decay model (see text).  The inset shows the hitting probability $p{\subs{hit}}(t_{\subs{hit}})/p_d$ as a function of $d$ with symbols from numerical simulations and the curve from the approximation of Eq. (\ref{phitapprox}) (see text).  }
\label{phit}
\end{center}
\end{figure}

The first question can be answered by calculating the average of
\begin{equation}
p_{\subs{hit}}(t) = |\langle \mbox{col} \ 2d | e^{-i \mathcal{H} t} |\mbox{col} \ 0\rangle|^2
\end{equation}
for instances of the SGT Hamiltonian with disorder strength $W$ and for various sizes $d$.  This corresponds to the single-shot measurement procedure \cite{Kempe03a} for the quantum walk, and a representative set of $p_{\subs{hit}}(t)$ as a function of time and values of $W$ are shown in Fig. \ref{phit}.  The first thing to observe is the oscillatory structure, due to the Bessel function structure for $p_{\subs{hit}}(t)$ \cite{Childs2002,Bose2003}, from which we can see that probability is maximized at the hitting time
\begin{equation}
t_{\subs{hit}} \approx \frac{1}{2\sqrt{2} \gamma} \left[ 2 d + 1 + 1.0188 \left(d + \frac{1}{2}\right)^{1/3} \right],
\end{equation}
with a value $p_d \sim d^{-2/3}$ for large $d$.  Second, as disorder increases the maximum of the hitting probability is seen to decrease.  This hitting probability can be approximated by 
\begin{equation}
p_{\subs{hit}}(t_{\subs{hit}}) \approx p_d \exp \left[ - \frac{1}{16} (d - \frac{1}{2}) W^2 \right],
\label{phitapprox}
\end{equation}
The exponential decay of $p_{\subs{hit}}/p_d$ is shown in the inset to Fig. \ref{phit}.  

\begin{figure}[b]
\begin{center}
\includegraphics[width=3.5in]{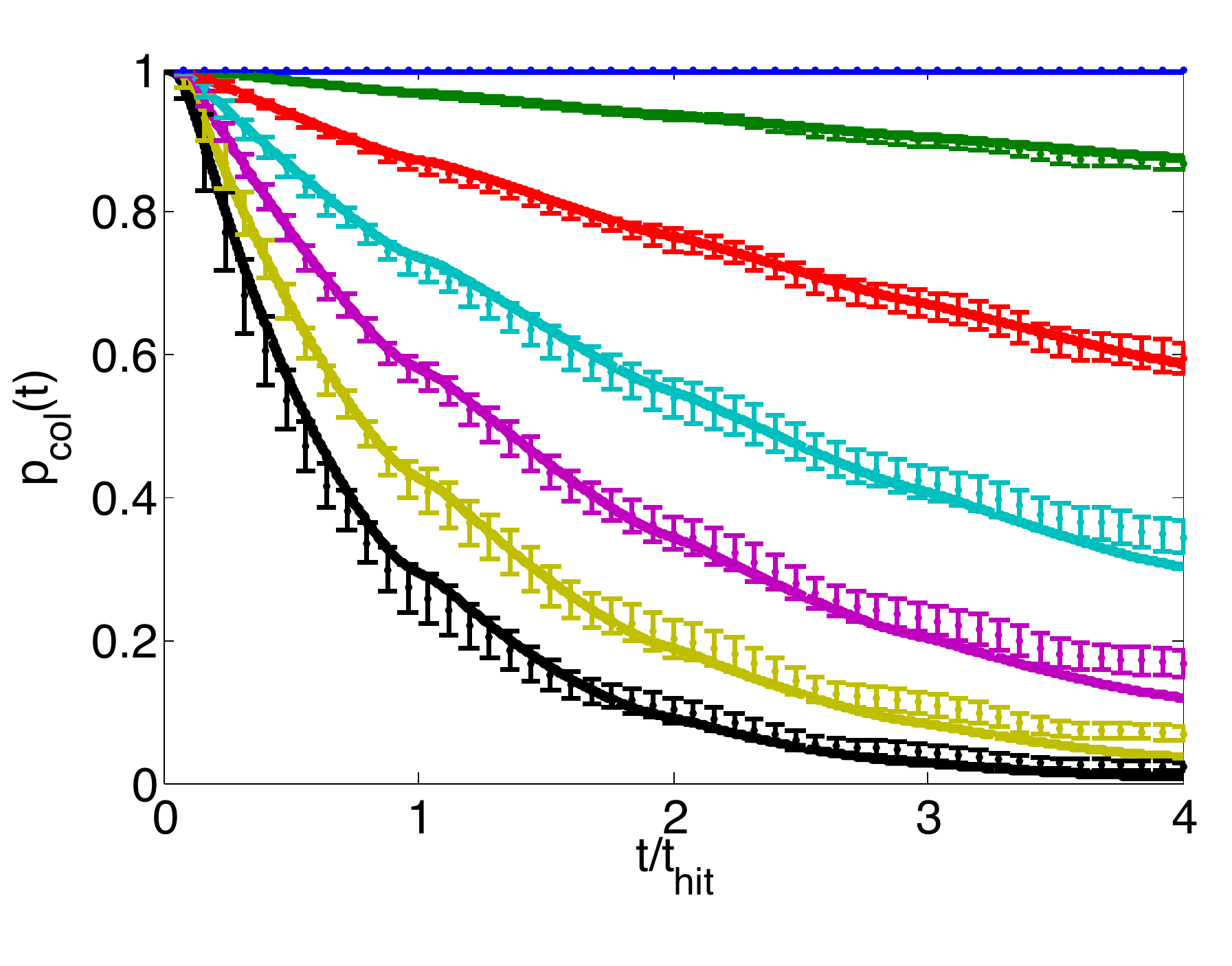}
\caption{(Color online) Column space probability $p_{\subs{col}}(t)$ as a function of time for the quantum walk on the glued trees graph with $d=15$ and various values of disorder.  The symbols were calculated using exact numerical simulations with (from top to bottom) $W = 0, 0.2, 0.4, 0.6, 0.8, 1.0, \mbox{and} \ 1.2$, with 10 realizations of $\mathcal{H}$ for each value of $W$.  The error bars indicate the standard deviation of each average.  The solid curves were calculated using the local decay model (see text). }
\label{pcol}
\end{center}
\end{figure}

By what mechanism does this decay occur?  It is not associated with eigenstate localization, as the eigenstates of the system are well within the extended regime in the phase diagram of Fig. \ref{IPRfig1}.  To understand this, we turn to another quantity of interest, the column-space probability
\begin{equation}
	p_{\subs{col}}(t) = \sum_{j=0}^{2d} \; \left| \,\braket{\mbox{col }j}{\psi(t)} \,\right|^2.
\end{equation}
Quantum walks on the glued trees graph that are initially in a column space state $| \mbox{col} \ j_0 \rangle$ will remain in the column space in the absence of disorder.   Once static disorder is introduced, the eigenstates lying within the column space become coupled to the other eigenstates of $G_d$.   These are associated with the subgraphs of $G_d$, and have zero amplitude on the graph's two roots.  The resulting decay of $p_{\subs{col}}(t)$, shown in Fig. \ref{pcol} leads to the decay of $p_{\subs{hit}}(t)$.  

We can provide an analytical estimate of this decay through perturbation theory.  Assuming the walk begins in a column state $\kc{j_0}$, taking the second order expansion of $p_{\subs{col}}(t)$ gives
\begin{eqnarray}
	p_{\subs{col}}(t) &=& \sum_{j=0}^{2d} \; \left| \,\bc{j} \exp(-i\,\ham\,t)\kc{j_0} \,\right|^2 \nonumber \\
	&\approx& \sum_{j=0}^{2d} \; |\braket{\mbox{col $j$}}{\mbox{col $j_0$}}|^2 + t^2 |\bc{j}\ham\kc{j_0}|^2 \nonumber \\
	& & \qquad  - t^2 \left( \braket{\mbox{col $j$}}{\mbox{col $j_0$}} \bc{j_0}\ham^2\kc{j} \right), \nonumber \\
\end{eqnarray}
where the Hamiltonian is given by $\ham = \ham_0 + \ham'$.  It is straightforward to show that $\ham_0$ contributes nothing to the quadratic term, and for $\ham'$ given by Eq. (\ref{hprime}) we find
\begin{equation}
	p_{\subs{col}}(t) = 1 - \frac{t^2}{N_{j_0,d}^2} \left[ (N_{j_0,d} - 1) \sum_i {\eps_i}^2
		- \sum_{i \ne j} \eps_i \, \eps_j \right].
\end{equation}
Averaging $p_{\subs{col}}(t)$ over the disorder we find
\begin{equation}
	\langle p_{col}(t) \rangle = 1 - \frac{1}{12} t^2 W^2 \left( 1 -\frac{1}{N_{j_0,d}} \right),
\end{equation}
where we have used $\langle \eps_i \eps_j \rangle = \frac{1}{12} W^2 \delta_{ij}$.  

This result for the short time decay applies to any graph for which the quantum walk can be projected onto a column space.  On longer timescales, we conjecture that the decay will have an exponential character, while retaining the position dependence. Hence, extrapolating from the short time result, we postulate a model of ``local (exponential) decay'', in which the walk evolution is computed as in the ideal column space representation, but the probability at each site $j$ is allowed to decay:
\begin{equation}
	p_{j}(t) = p_0 \exp\left[ - \frac{t}{12} \frac{W^2}{\gamma} \left(1-\frac{1}{N_{j,d}} \right) \right]. \label{loc_gamma}
\end{equation}
This is equivalent to applying the mapping 
\begin{equation}
	\ham \quad\mapsto \quad \hcol - \, i\Gamma/2, \label{ldm}
\end{equation}
where $\hcol$ is the projection of the graph Hamiltonian onto the column space, and $\Gamma$ is given by
\begin{equation}
	\Gamma = \frac{1}{12} \frac{W^2}{\gamma}\, \sum_{j=0}^{2d}\left(1-\frac{1}{N_{j,d}} \right) \kc{j}\bc{j},
\end{equation}
with $\gamma$ denoting the unit time hopping probability from $\mathcal{H}_0$.  This expression for $\Gamma$ can be anticipated by applying Fermi's Golden Rule to $\mathcal{H}'$ in the eigenstate basis, but the presence of a large discrete component to the spectrum ($\sigma_0$) prevents us from making a precise derivation.  Numerical evidence, however, suggests that this is in fact the correct mechanism.

Along with the exact numerical results for $p_{\subs{hit}}$ and $p_{\subs{col}}$ in Figs. \ref{phit} and \ref{pcol}, we have included results from the local-decay model calculated in the (exponentially smaller) column-space representation for $\mathcal{H}_0$.  We observe that the local decay model predictions (solid lines) agree well with the results of simulations (points), to within one standard deviation.   Our model accurately reproduces many features of  the exact simulations, such as the variation in the decay rate of $p_{\subs{col}}$ and the oscillations in $p_{\subs{hit}}$.  Furthermore, if we use the column space probability to predict the hitting probability at the target node (opposite root), the agreement is quite good.  This supports the claim that the disorder-induced reduction in quantum transport is primarily explained by decay from the column space.  

\begin{figure}
\begin{center}
 \includegraphics[width = 3.5in]{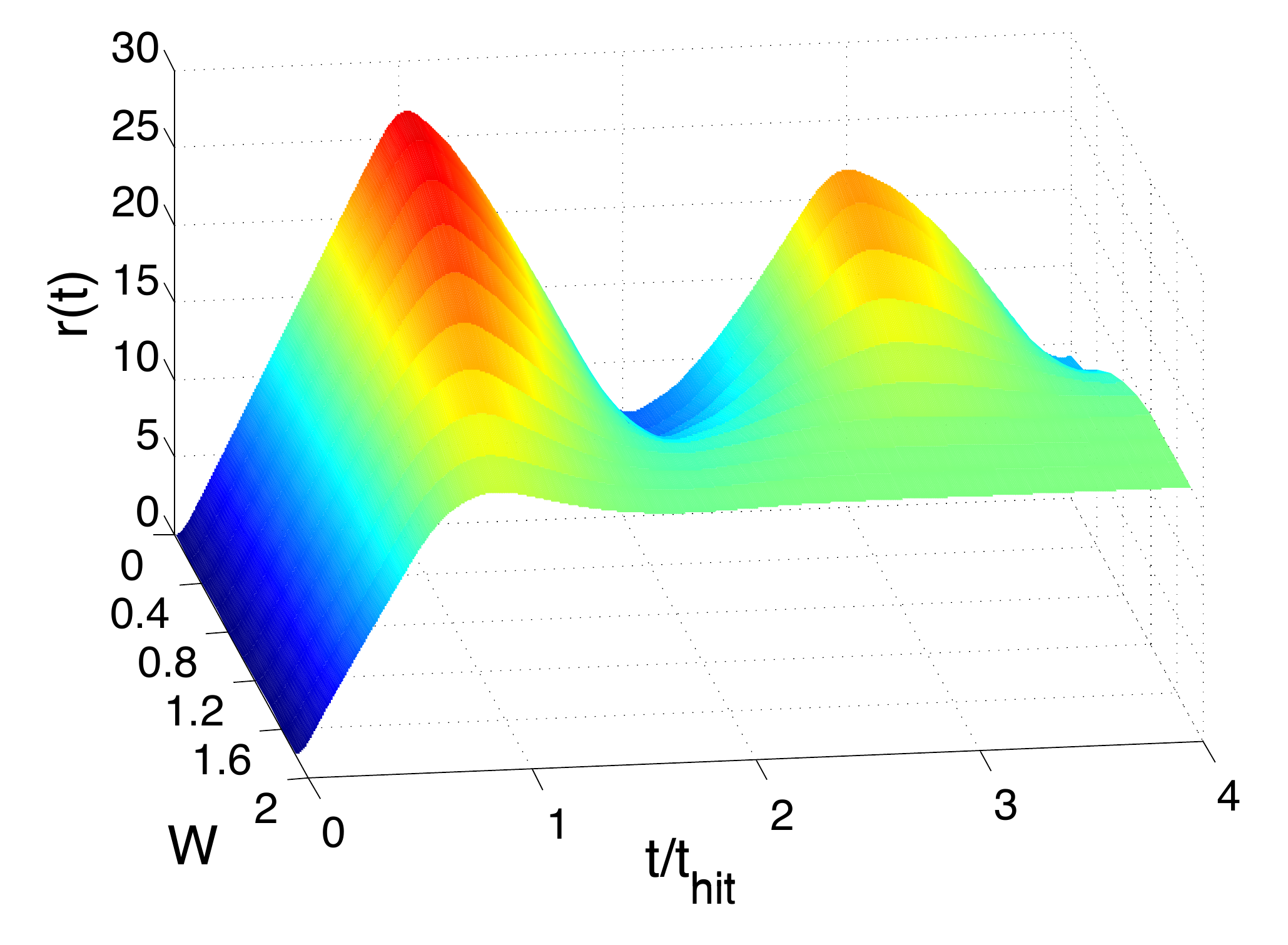}
\caption{(Color online) Average distance $r(t)$ for $d=15$ as a function of time and disorder $W$.  For each $W$, the simulations were averaged over $10$ realizations of $\mathcal{H}$.  The quantum oscillations decay in time for small disorder, with a critical damping near $W \approx 2$, indicating a type of quantum-to-classical transition.}
\label{rtime}
\end{center}
\end{figure}

So far we have looked at the probability at the right root and the column space, but a more global characterization of the walk propagation can be found by analyzing the average depth reached by the quantum walk,
\begin{equation}
	r(t) = \bra{\psi(t)} \;\hat{r}\; \ket{\psi(t)},
\end{equation}
where the column position operator $\hat{r}$ is defined as
\begin{equation}
\hat{r} = \sum_{j=0}^{2d} \sum_{n=0}^{N_{j,d}-1} j |j,n\rangle \langle j,n|
\end{equation}
with the property $\hat{r} \kc{j} = j \kc{j}$.  The value of $r(t)$ thus gives a snapshot of the expected position of the walk along the graph at any one time.  This is displayed in Fig. \ref{rtime} as a function of time and disorder.  In the absence of disorder, the average depth shows an oscillatory character consistent with ballistic propagation of the wavepacket and reflections at the two ends of the graph.  Disorder-induced decay from the column space causes the amplitude of the oscillations to decay faster than in the ideal case, ultimately causing a ``damping'' of the oscillations. Therefore, the walk has a reduced probability of traversing the graph, and substantial probability is instead deposited in the center of the graph, where the concentration of nodes is highest.

\begin{figure}
\begin{center}
\includegraphics[width = 3.5in]{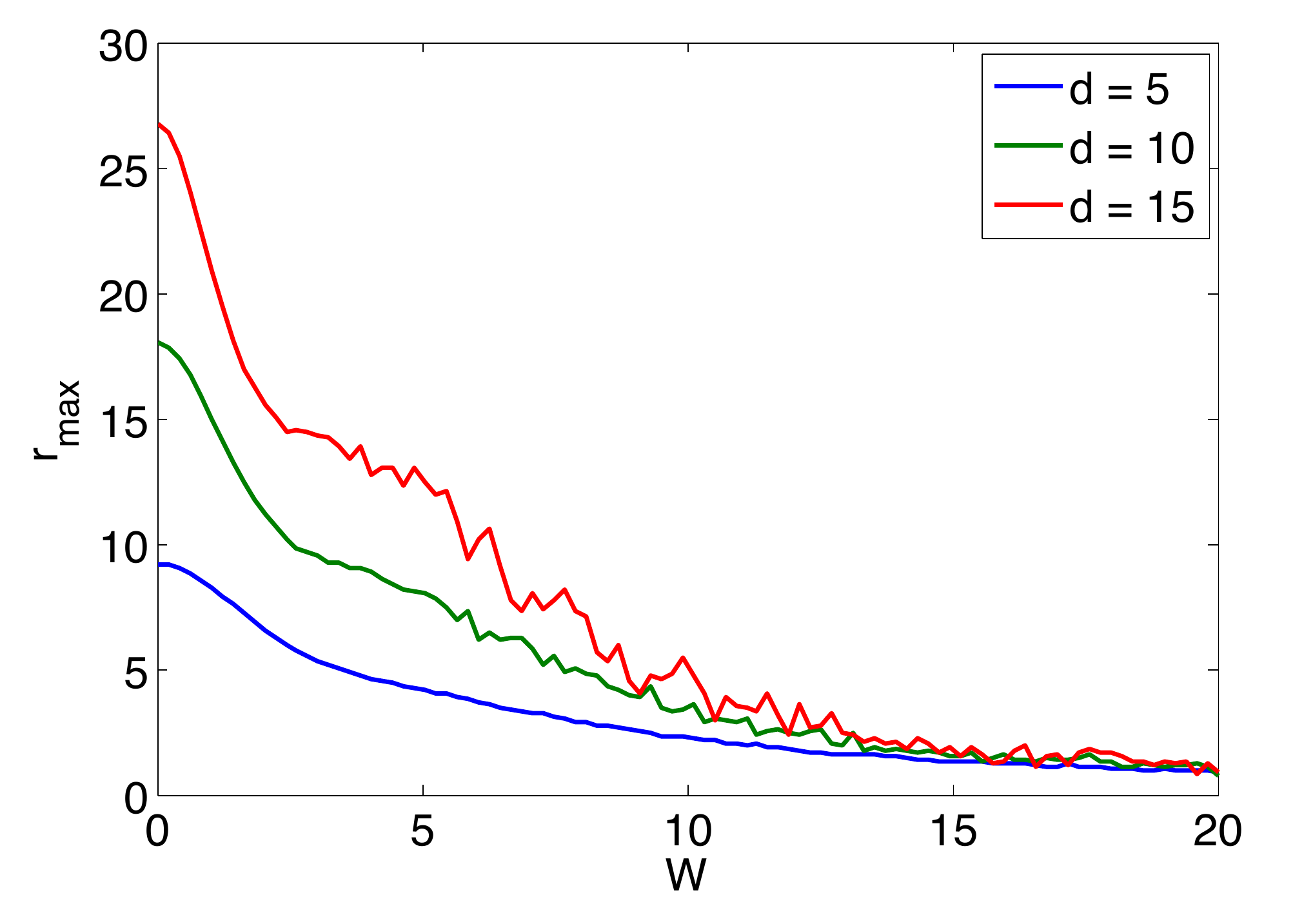}
\caption{(Color online) Maximum average distance $r(t)$ as a function of disorder $W$ for various depths.  For each value of disorder the maximum of $r(t)$ was averaged over $1000, 100$, and $10$ realizations for $d=5, 10$, and $15$, respectively.   For intermediate disorder ($2 < W < 4$), there is a quantum-to-classical transition, while for large disorder ($W>15$) a localization transition is observed.}
\end{center}
\label{rmax}
\end{figure}

In Fig. \ref{rmax} we plot the maximum value of $r(t)$ in the range $t < 3 t_{\subs{hit}}$ against the strength of disorder, illustrating the localization transition on the glued trees graph.  We see that for small disorder ($W < 2$), the maximum value of $r$ is high, corresponding to the quantum walk hitting the right root, for which $r = 2d$.  As disorder increases, the curve falls, and begins to level off at the graph center ($r \approx d$ for $2 < W < 4$). For larger amounts of disorder ($W > 4$), the curves continue to decrease, converging on a single value around $W = 16$, precisely when we expect all eigenstates to be localized.

This analysis of the quantum walk dynamics strongly suggests a type of quantum-to-classical or ``wavelike-to-diffusive'' crossover at weak disorder. Note that this not a ``ballistic-to-diffusive'' crossover, and does not conflict the ballistic spreading of wavepackets found by Klein \cite{Klein95} for the Bethe lattice at weak disorder, as classical diffusion also leads to ballistic spreading ($ r(t) \sim t$).  It does, however, suggest that there may be a length scale (the mean-free-path) that limits the size of graphs for which a speedup could occur in the presence of disorder.  That is, the exponential decay of the hitting probability (in $d$) seen in Fig. \ref{phit} may be interpreted as the classical probability for a walker to traverse $d$ sites given a mean-free-path of order $\ell \sim 1/W^2$.   To understand this crossover in more detail we extend our analysis of the quantum walk to a transport model.  

\section{Quantum Walk Transport}

\begin{figure}
\begin{center}
 \includegraphics[width = 3.5in]{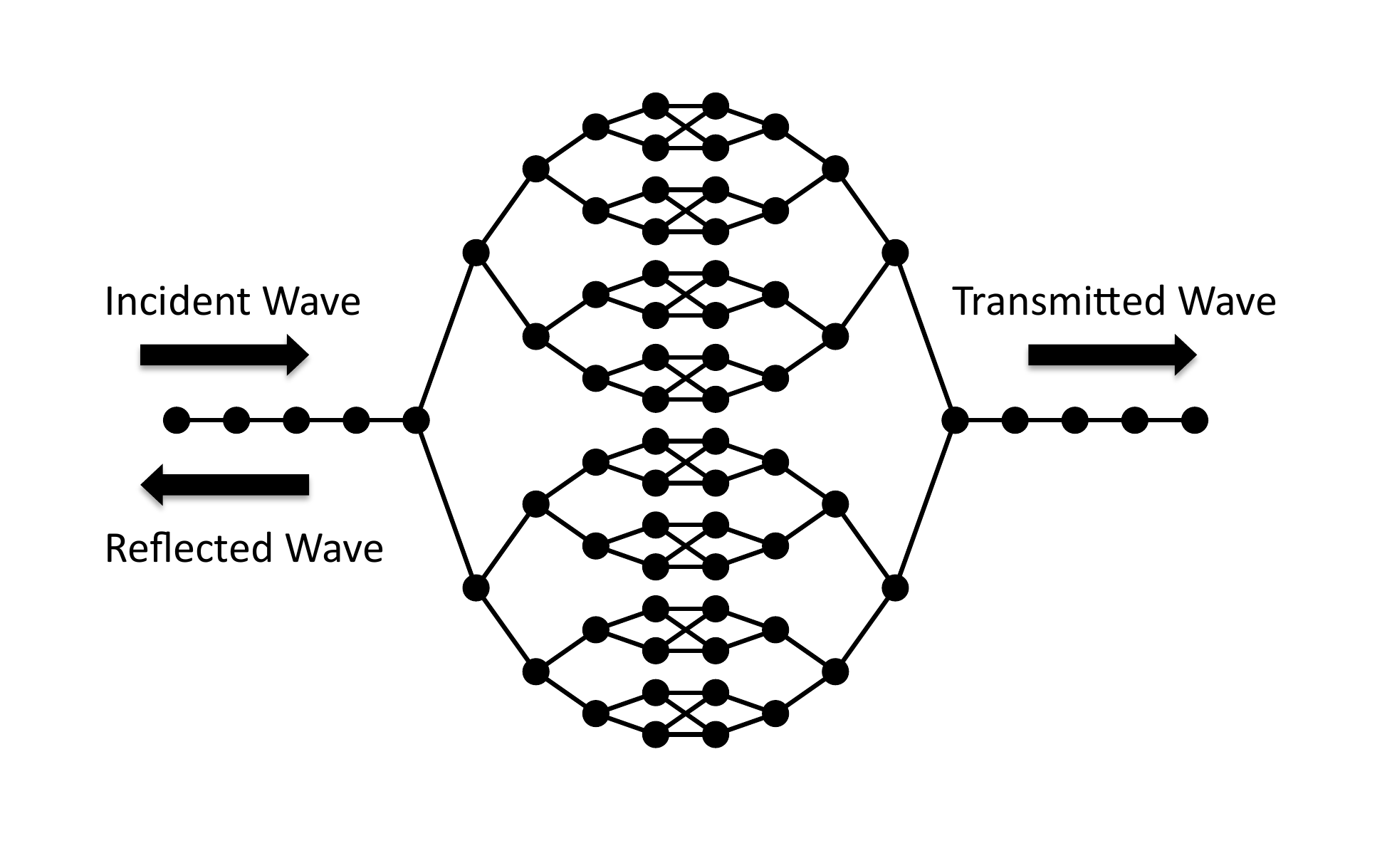}
\caption{Scattering approach to the quantum walk, in which an incident wavepacket is transmitted and reflected along the ``tails'' connected to the modified glued trees graph.}
\label{scatterfig}
\end{center}
\end{figure}

To identify the quantum-to-classical transition, we consider the quantum transmission through the glued trees graph, subject to disorder.  To transform the quantum walk into a transmission problem, we attach ``tails'' to the input and output nodes, and look at the transmission coefficient through the graph for a wavefunction of the form
\begin{equation}
\Psi(n) = \left\{ \begin{array}{ll} 
e^{i k n} + \mathcal{R} e^{-i k n} & \mbox{for} \ n < 0 \\
\mathcal{T} e^{i k n} & \mbox{for} \ n > 2d+1 \end{array} \right.
\end{equation}
This represents an ingoing wave that is reflected and transmitted through the graph, as illustrated in Fig. \ref{scatterfig}.  This type of quantum walk was been used to develop a quantum algorithm for NAND-tree evaluation \cite{Farhi2008}, and has been generally analyzed in \cite{Varbonov2009}.  We consider the transmission probability $T = |\mathcal{T}|^2$ as a function of depth and disorder. 

Using a standard analysis for transmission in tight-binding lattices \cite{Sadreev2003}, we find the transmission amplitude
\begin{equation}
\mathcal{T} = \langle \mbox{col} \ 0 | \frac{2 i \sin k}{\tilde{\mathcal{H}} - 2 \cos k} | \mbox{col} \ 2 d+1\rangle
\end{equation}
where
\begin{equation}
\tilde{\mathcal{H}} = \mathcal{H} + e^{i k} \left( | \mbox{col} \ 0\rangle \langle \mbox{col} \ 0| + |\mbox{col} \ 2d+1 \rangle \langle \mbox{col} \ 2d+1| \right)
\end{equation}
is an effective Hamiltonian for the MGT graph alone (note that this graph has $2d+1$ columns).  This quantity can be calculated by diagonalizing the non-Hermitian Hamiltonian $\tilde{\mathcal{H}}$, and forming the appropriate matrix elements in $\mathcal{T}$.  

\begin{figure}
\begin{center}
 \includegraphics[width = 3.5in]{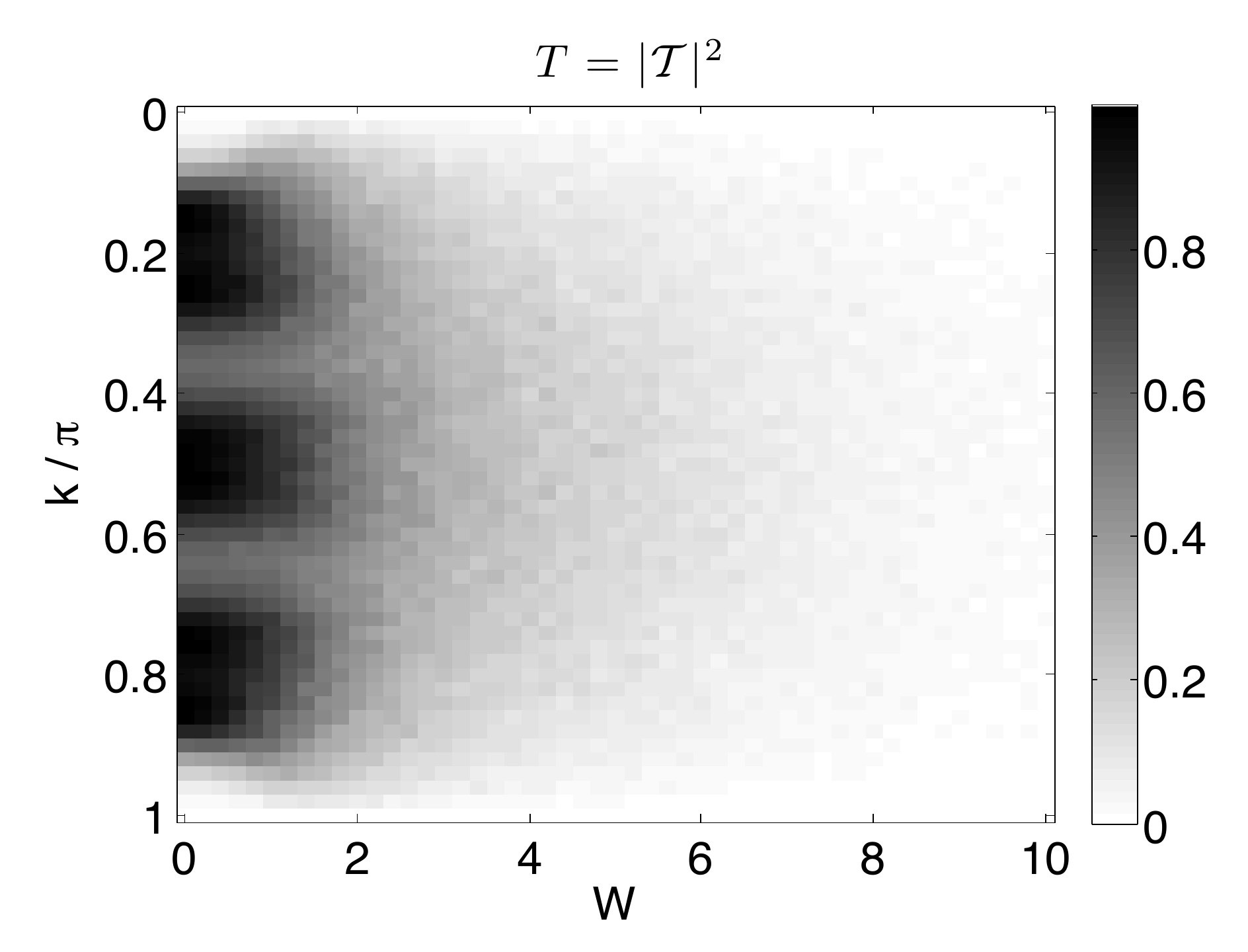}
\caption{Transmission probability $T$ as a function of momentum $k$ and disorder $W$ for a modified glued trees graph of depth $d=5$.  For each value of momentum and disorder, the transmission was averaged over 250 realizations of $\tilde{\mathcal{H}}$. }
\label{transfig1}
\end{center}
\end{figure}

\begin{figure}
\begin{center}
 \includegraphics[width = 3.5in]{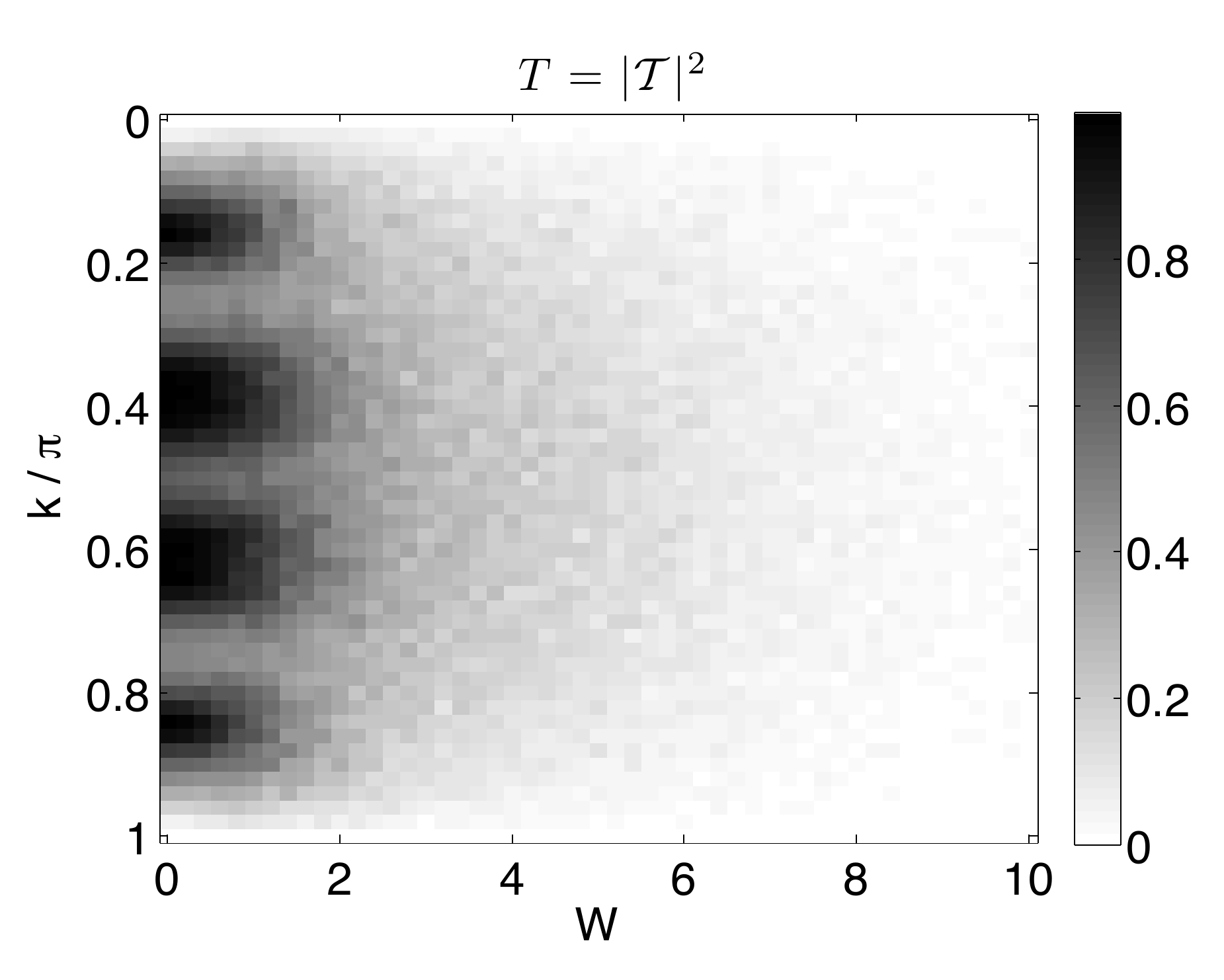}
 \caption{Transmission probability $T$ as a function of momentum $k$ and disorder $W$ for a modified glued trees graph of depth $d=6$. For each value of momentum and disorder, the transmission was averaged over 100 realizations of $\tilde{\mathcal{H}}$.}
\label{transd6}
\end{center}
\end{figure}

\begin{figure}
\begin{center}
 \includegraphics[width = 3.5in]{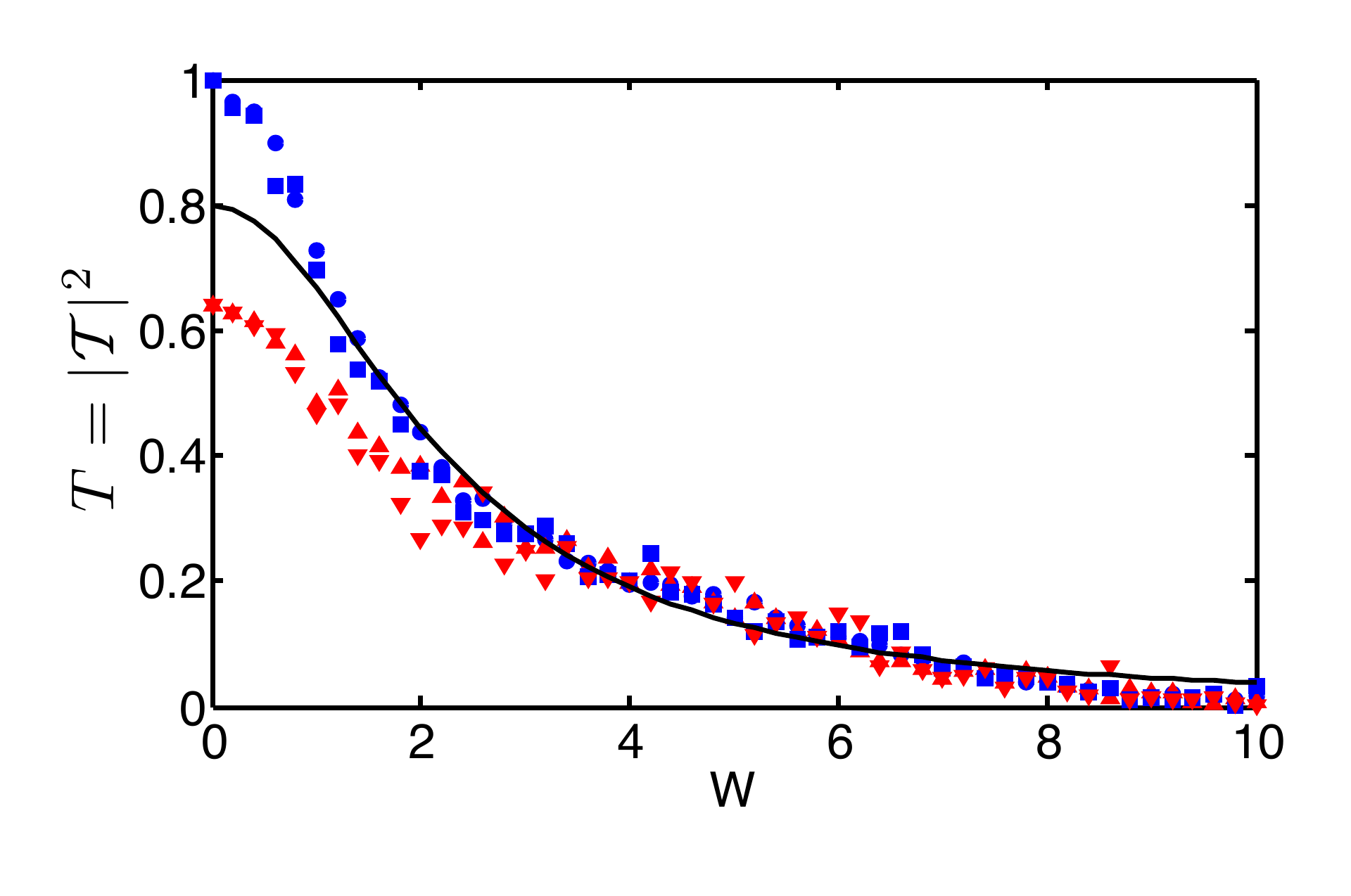}
\caption{(Color online) Transmission probability $T$ for $k=\pi/2$ as function of disorder and for various depths.  The various symbols are $d = 7$ (blue circles),  $d=8$ (red upward triangles), $d=9$ (blue squares), and $d=10$ (red downward triangles), averaged over $500, 200, 100, \mbox{and} \  50$ realizations of $\tilde{\mathcal{H}}$, respectively.  The solid black curve is the transmission probability for the classical random walk $T_c$ (see text).}
\label{transmission_fit}
\end{center}
\end{figure}

The resulting transmission probability $T = |\mathcal{T}|^2$ is shown as a function of momentum $k$ and disorder $W$ in Figs. \ref{transfig1} and \ref{transd6}, for depths $d=5$ and $d=6$, respectively.  For small disorder, there is a size-dependent oscillatory structure as a function of $k$ due to resonances, much like those found in \cite{Sadreev2003} (and briefly described in Appendix A).   These oscillations in $T$ disappear when the disorder strength $W \approx 2$, after which the transmission decays monotonically.   Figure \ref{transmission_fit} shows $T$ for $k=\pi/2$ as a function of disorder for many graph sizes, which all exhibit the same behavior for $W > 2$, indicative of a transition in $T$.    

The decay of the transmission probability with disorder can be understood using a classical model, described in the Appendix B.  This model uses a diffusion constant proportional to the mean-free-path $\lambda \sim \ell \sim W^{-2}$ and leads to a classical transmission probability of the form
\begin{equation}
T_c = \frac{T_0}{1 + c (W/\gamma)^2},
\label{T_c}
\end{equation}
where the coefficients $T_0$ and $c$ presumably depend on the exact mapping of the disordered quantum walk to the diffusion equation, such as the method of \cite{Amir2009} or the results of \cite{Erdos2005}.   This expression for Eq. (\ref{T_c}), fit using $T_0 = 0.8$ and $c = 0.2$, is shown in Fig. \ref{transmission_fit}.  This agreement, for intermediate values of disorder, provides confirmation of the quantum-to-classical crossover observed in the dynamical studies of the previous section.    

\section{Conclusion}

In this paper, we have carried out an investigation of the effects of diagonal disorder on quantum walks on the glued trees graph.  While disorder does lead to localization in the strong disorder limit, we find the primary effect in the case of small disorder to be quantum decay out of the column space. The quantum decay can be accurately modeled in the column space by a non-unitary mapping that enforces position dependent decay of the probability. This local decay model is efficient to compute, owing to the exponential reduction in size of the representation, yet it allows prediction of the end-to-end hitting probability, and should be extendable to other graphs with similar symmetries.

One such graph is the hypercube.  This problem had been previously studied for quantum state transfer \cite{Strauch2008}, where the effects of off-diagonal disorder were emphasized.   Numerical simulations for diagonal disorder, however, provide very similar results to those found in Sec. IV, with one important difference.  The hitting time for the hypercube is independent of the dimension $d$ (which is analogous to the depth of the glued binary trees graph).  The local decay model then predicts that the hitting probability should decay as $e^{-W^2/12}$, a result borne out by simulations.  Thus, for the hypercube, quantum transport outperforms classical transport for small $W$.  

Such a result is also possible for the glued trees graph.   The exponential suppression of the hitting probability occurs for the specific case of a state initially localized to the left root of the graph.  By using a graph with ``tails'', the results of Sec. V show that appreciable transport is possible for $W<2$, provided there is no disorder in the tails and an appropriate initial state can be found.  In addition, alternative measurement strategies \cite{Varbonov2008}, or the inclusion of traps \cite{Mulken2011} could provide opportunities for speedup.  Exploring this possibility could provide additional context for understanding environmentally assisted quantum transport \cite{AGuzik08}.

In summary, we have performed an analysis of the effect of static disorder on a quantum walk on the glued trees graph.  For small disorder, we find that the dominant effect is a type of quantum decay, and not quantum localization.  For intermediate disorder, there is a crossover to diffusive transport, while a localization transition is observed at large disorder, in agreement with Anderson localization on the Cayley tree.  Our results suggest that intermediate disorder will inhibit any quantum speedup, but also that large speedups are possible for quantum walks on complex networks with small disorder.  

\begin{acknowledgments}
We thank A. Aspuru-Guzik, S. M. Girvin, T. Kottos, and S. Lloyd for helpful discussions.  FWS was supported by the Research Corporation for Science Advancement.
\end{acknowledgments}

\appendix

\section{Quantum Walk Transport}

The transmission through the modified glued trees graph can be calculated by using the following ansatz for the column-space wavefunction
\begin{equation}
\Psi_{\subs{MGT}}(n) = \left\{ \begin{array}{ll} 
e^{i k n} + \mathcal{R} e^{-i k n} & \mbox{for} \ n < 0, \\
A e^{i \tilde{k} n} + B e^{-i \tilde{k} n} & \mbox{for}  \ 0 < n < d, \\
C e^{i \tilde{k} n} + D e^{-i \tilde{k} n} & \mbox{for} \ d+1 < n < 2d+1,\\
\mathcal{T} e^{i k n} & \mbox{for} \ n > 2d+1. \end{array} \right. 
\end{equation}
To determine $\mathcal{T}$, one must use continuity of $\Psi$ at $n=0$ and $2d+1$, and the Schr{\"o}dinger equation at $n = 0, d, d+1, $ and $2d+1$, with $E = - 2 \gamma \cos k = -2 \sqrt{2} \gamma \cos \tilde{k}$.  These provide six equations for the six unknowns $A, B, C, D, \mathcal{R},$ and $\mathcal{T}$.  For general $k$, these equations are most conveniently solved by computer, and have a rather complicated solution.  For $k=\pi/2$, however, the solution for $\mathcal{T}$ is remarkably simple
\begin{equation}
\mathcal{T}(k=\pi/2) = \frac{8}{9 + (-1)^d}.
\end{equation}
This oscillation of $\mathcal{T}(k=\pi/2)$ as a function of $d$ for the MGT facilitated the classical interpretation of Fig. 12.

For comparison, the appropriate ansatz for the simple glued trees graph is
\begin{equation}
\Psi_{\subs{SGT}}(n) = \left\{ \begin{array}{ll} 
e^{i k n} + \mathcal{R} e^{-i k n} & \mbox{for} \ n < 0, \\
A e^{i \tilde{k} n} + B e^{-i \tilde{k} n} & \mbox{for}  \ 0 < n < 2d, \\
\mathcal{T} e^{i k n} & \mbox{for} \ n > 2d, \end{array} \right.
\end{equation}
and the corresponding solution for the transmission coefficient is  $\mathcal{T}(k=\pi/2) = 1$, independent of $d$.

\section{Classical Walk Transport}

\begin{figure}
\begin{center}
 \includegraphics[width = 3.5in]{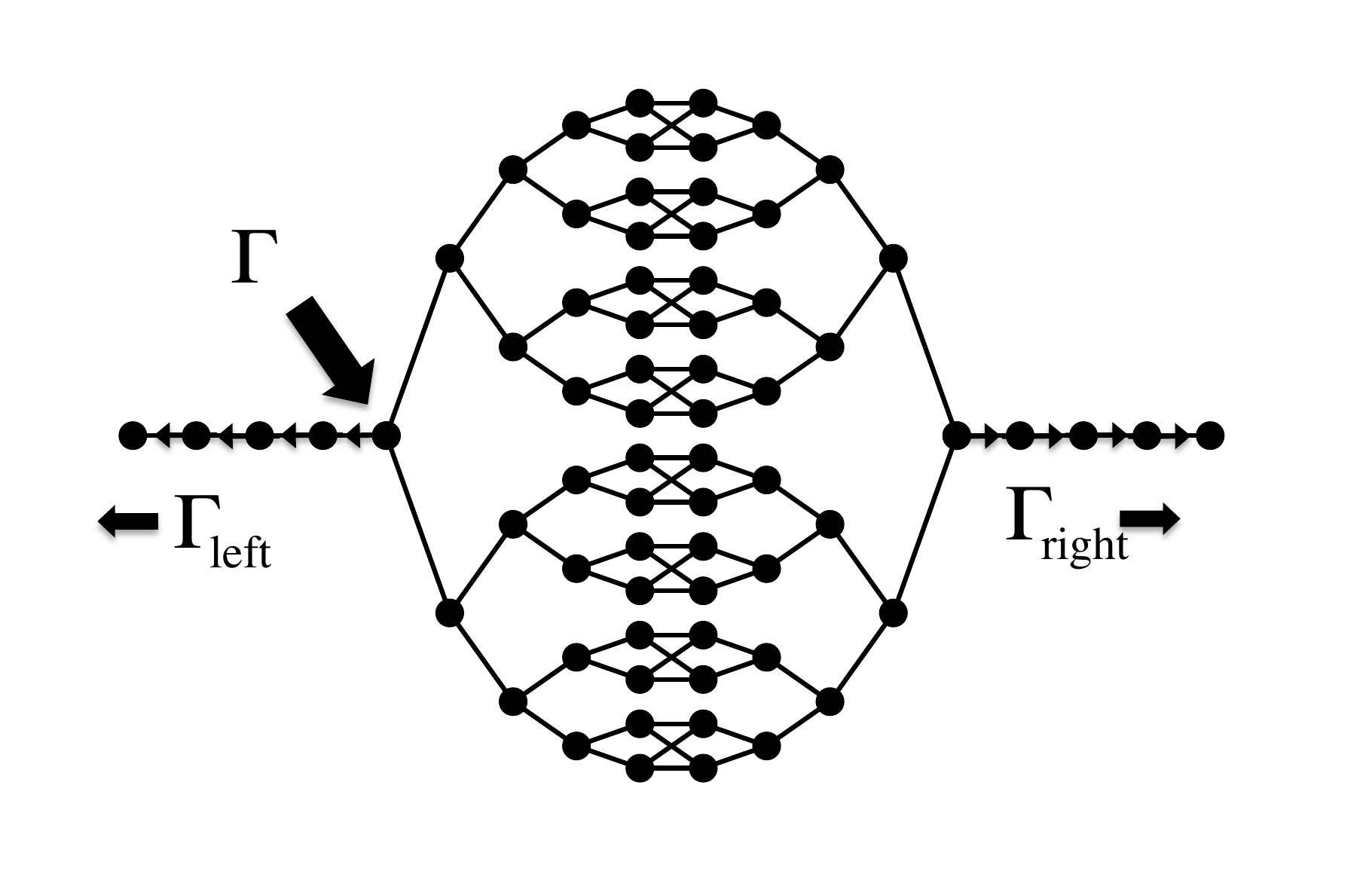}
\caption{Transport approach to classical random walk on the modified glued trees graph.  A steady-state flux is $\Gamma$ introduced to the left root and carried away by the ``tails'' on the left and the right, with $\Gamma = \Gamma_{\subs{left}} + \Gamma_{\subs{right}}$. }
\label{classicalfig}
\end{center}
\end{figure}

The scattering theory approach of Sec. IV can be modified to study classical diffusive transport through the modified glued trees graph.  The model, depicted in Fig. \ref{classicalfig}, consists of an ingoing particle flux $\Gamma$ to the left root.  These particles then undergo a random walk in the interior of the graph, with diffusion rate $\lambda$, but can escape from either the left root (with rate $\lambda_{\subs{left}}$) or the right root (with rate $\lambda_{\subs{right}}$).  This is indicated by the directed paths on the left and right ``tails'' on the graph.  In the steady-state, there is an outgoing flux to the left $\Gamma_{\subs{left}} = p_{\subs{left}} \lambda_{\subs{left}}$ and an outgoing flux to the right $\Gamma_{\subs{right}} = p_{\subs{right}} \lambda_{\subs{right}}$, with $\Gamma = \Gamma_{\subs{left}} + \Gamma_{\subs{right}}$.   The classical transmission coefficient introduced in the text is $T_c = \Gamma_{\subs{right}}/\Gamma$.  

The calculation of $T_c$ can be found by solving for the steady-state of the master equation
\begin{equation}
\frac{d p_j}{dt} = \sum_k L_{j,k} p_k  + b_j,
\end{equation}
where $L_{j,k}$ is a matrix of rate constants, and $b_j$ is zero except for the input site (where it equals $\Gamma$).  Due to the symmetry of the modified glued trees graph, we can write this equation in terms of the probabilities for the walker to be on each column, so that $j=0 \to 2d+1$, $p_{\subs{left}} = p_0$, $p_{\subs{right}} = p_{2d}$, and $b_j = \Gamma \delta_{j,0}$.  One can find the steady-state probability distribution from $\vec{p}_{\subs{ss}} = - L^{-1} \vec{b}$. 

Here we provide an analytical calculation by solving for the steady state by recursion relations.  In full, the master equation reads
\begin{equation}
\frac{dp_0}{dt} = - (2\lambda + \lambda_{\subs{left}}) p_0 + \lambda p_1 + \Gamma,
\end{equation}
\begin{equation}
\frac{dp_j}{dt} = \left\{ \begin{array}{ll}
2 \lambda p_{j-1} - 3 \lambda p_{j} + \lambda p_{j+1} & \mbox{for} \ 0< j < d \\
2 \lambda p_{j-1} - 3 \lambda p_j + 2 \lambda p_{j+1} & \mbox{for}\  j=d \ \mbox{and} \ d+1 \\
\lambda p_{j-1} - 3 \lambda p_{j} + 2 \lambda p_{j+1} & \mbox{for} \ d+1 < j < 2 d+1 \\
\end{array} \right.
\end{equation}
and
\begin{equation}
\frac{dp_{2d+1}}{dt} = \lambda p_{2d} - (2\lambda + \lambda_{\subs{right}}) p_{2d+1}.
\end{equation}
For the steady-state $d \vec{p}/dt = 0$, we can solve the reccurence relations
\begin{eqnarray}
p_{j+1} &=& 3 p_{j} - 2 p_{j-1} \quad \mbox{for} \ j = 1 \to d \\
p_{j-1} &=&  3 p_{j} - 2 p_{j+1} \quad \mbox{for} \ j = 2d+1 \to d
\end{eqnarray}
in terms of $p_0$ and $p_{2d+1}$, respectively.  We find
\begin{eqnarray}
p_j &=& 2^j p_0 + (2^j - 1) \frac{\lambda_{\subs{left}}}{\lambda}  p_0 - (2^j-1) \frac{\Gamma}{\lambda} \\
p_{2d-j} &=& 2^j p_{2d+1} + (2^j-1) \frac{\lambda_{\subs{right}}}{\lambda} p_{2d+1}
\end{eqnarray}

Requiring these two to solve $dp_d/dt=0$, we obtain one equation for $p_0$ and $p_{2d+1}$:
\begin{equation}
\begin{array}{c}
\left[2^d + (2^d-1) \frac{\lambda_{\subs{left}}}{\lambda}\right] p_0 - \left[2^d + (2^d-2^{-1}) \frac{\lambda_{\subs{right}}}{\lambda}\right] p_{2d+1} \\
\quad \quad \quad = (2^d-1) \frac{\Gamma}{\lambda}.
\end{array}
\end{equation}
This can be combined with the conservation law
\begin{equation}
\lambda_{\subs{left}} p_0 + \lambda_{\subs{right}} p_{2d+1} = \Gamma,
\label{masterc2}
\end{equation}
to solve for $p_0$ and $p_{2d+1}$.  For the latter, we find
\begin{equation}
p_{2d+1} = \frac{ \Gamma / \lambda_{\subs{left}}}{1 +\left(\frac{\lambda_{\subs{right}}}{\lambda_{\subs{left}}}\right) + 2 (1-3 \times 2^{-d-2}) \left(\frac{\lambda_{\subs{right}}}{\lambda}\right)}.
\end{equation}
Forming $T_c = \lambda_{\subs{right}} p_{2d+1} / \Gamma$, and simplifying we find
\begin{equation}
T_c = \frac{1}{1 +\left(\frac{\lambda_{\subs{left}}}{\lambda_{\subs{right}}}\right) + 2 (1- 3 \times 2^{-d-2}) \left(\frac{\lambda_{\subs{left}}}{\lambda}\right)}.
\end{equation}

To complete our calculation, we estimate the diffusion rate $\lambda$ as the product of the hopping rate ($\gamma$) and the transition probability between sites ($\gamma^2/W^2$, for large $W$).  Finally, setting $\lambda_{\subs{left}} = \lambda_{\subs{right}} = \gamma$ and $\lambda = \gamma^3/W^2$ yields 
\begin{equation}
T_c = \frac{1/2}{1 + (1-3 \times 2^{-d-2}) (W/\gamma)^2},
\end{equation}
a result very similar to that used in Sec. VI.

\bibliography{qwalk}

\end{document}